\documentstyle[preprint,aps]{revtex}
\begin{document}
\draft
\preprint{CLNS-95/1367}
\title{Radiative Decays of Excited $\Lambda_Q$ Baryons \\
in the Bound State Picture}
\author{Chi-Keung Chow}
\address{Newman Laboratory of Nuclear Studies, Cornell University, Ithaca,
NY 14853.}
\date{\today}
\maketitle
\begin{abstract}
It is shown that, in the bound state picture, the $\Lambda_c(2593)
\to\Lambda_c\gamma$ and $\Lambda_c(2625)\to\Lambda_c\gamma$ decays are
severely suppressed.
On the other hand, for their bottom counterparts, which are predicted to have
masses 5900 and 5926 MeV respectively, may have significant radiative
branching ratio.
In particular, the $\Lambda_b(5926)\to\Lambda_b\gamma$ mode may even
dominate over the strong decay mode.
The isospin allowed $\Lambda^{**}_Q\to\Sigma_Q\gamma$ mode is expected to be
small.
\end{abstract}
\narrowtext
In Ref.~\cite{5,6,7,8}, the bound state picture of heavy baryons was
developed.
This model is motivated by the large $N_c$ limit, in which the ground state
light baryons (nucleon N and Delta $\Delta$) just appear as topological
solitons of the Goldstone fields.
As a result, the interactions of light baryons and heavy mesons can be
studied under chiral perturbation theory.
It turns out that, under chiral SU(2)$_L\;\times\;$SU(2)$_R$, the only
stable bound states are those with $K=I+s_\ell=0$.
These states can be identified with the low-lying heavy baryons $\Lambda_Q$
and $\Sigma^{(*)}_Q$, with $I=s_\ell=0$ and 1 respectively.
Then one can deduce the decay properties of the bound state from those of
the constituents.
For example, the $\Sigma_c\to\Lambda_c\pi$ decay was studied in Ref.~\cite{5}.
The decay rate is determined by the axial current coupling $g_3$ defined in
Ref.~\cite{9}.
\begin{equation}
\Gamma(\Sigma_c^{(*)}\to\Lambda_c\pi)={g_3^2\over6\pi}
{|\vec p_\pi|^3\over f^2}.
\end{equation}
In Ref.~\cite{5}, it was shown that
\begin{equation}
g_3=g_{soliton}+g_{heavy}
\end{equation}
where
\begin{mathletters}
\begin{equation}
g_{soliton}=\sqrt{\textstyle{3\over2}}g_A,
\end{equation}
\begin{equation}
g_{heavy}=-\sqrt{\textstyle{1\over6}}g.
\end{equation}
\end{mathletters}
Here $g_A=1.25$ is the nucleon axial current coupling and $g$ is the heavy
meson axial current coupling constant defined in Ref.~\cite{10,11,12}.
It is also possible to investigate other decay modes like
$\Sigma_c\to\Lambda_c\gamma$ \cite{8.1} in this picture.

In this paper, we are going to study the electric dipole decay of the
orbitally excited $\Lambda_Q$, denoted by $\Lambda^{**}_Q$
\footnote{There is no universal nomenclature for these excited
$\Lambda_c$ baryons.
In Ref.~\cite{13} they are denoted by $\Lambda^{(*)}_{c1}$,
while J. K\"orner call them $\Lambda^{**}_c$ in analogy to $D^{**}$ in the
meson sector.
The experimentists usually just call them by their masses, i.e., $\Lambda_c
(2593)$ and $\Lambda_c(2625)$.
The choice of nomenclature in this article just reflects the preference of
the author.} in this article.
The $\Lambda_c(2625)$ and $\Lambda_c(2593)$ doublet, seen by ARGUS\cite{8.2},
CLEO\cite{8.3} and E687\cite{8.4}, decay dominantly through the channels
$\Lambda_c(2625)\to\Lambda_c\pi\pi$ and $\Lambda_c(2593)\to\Sigma_c\pi$.
The absence pf the $\Lambda_c\pi$ mode indicates that these are $I=0$ states,
i.e., excited $\Lambda_c$'s rather than $\Sigma_c$'s.
Theoretically, these states have been studied in Ref.~\cite{13} under chiral
perturbation theory and heavy quark symmetry.
It is noted that the $\Lambda_c(2593)\to\Sigma_c\pi$ channel is severely
suppressed by phase space ($M_{\Lambda_c(2593)}-M_{\Sigma_c}=141$ MeV).
The isospin allowed $\Lambda_c(2625)\to\Sigma^*_c\pi$ channel is simply
kinematically forbidden ($M_{\Lambda_c(2625)}-M_{\Sigma^*_c}=105$ MeV) and
have to go through a three-body decay.
As a result, it is possible that their electromagnetic decay to $\Lambda_c$ or
$\Sigma^{(*)}_c$ may have a significant branching ratio.
However, there are undetermined coupling constants in the lagrangian in
Ref.~\cite{13} and no definite predictions are given on the actual decay
widths.

If these excited $\Lambda_Q$'s are P-wave states as commonly believed, their
parities will be opposite to that of the ground states $\Lambda_Q$ and
$\Sigma_Q$.
As a result, they can decay through electric dipole transitions to these
ground states.
The decay width is given by
\begin{equation}
\Gamma(\Lambda^{**}_Q\to\Lambda_Q\gamma)=\textstyle{4\over3}|\vec p_\gamma|^3
|\vec{\bf P}|^2,
\end{equation}
where $\vec{\bf P}$ is the electric dipole moment between the initial and
final state.
This width has the order of magnitude of
\begin{equation}
\Gamma(\Lambda^{**}_Q\to\Lambda_Q\gamma)\sim|\vec p_\gamma|^3 \alpha r^2,
\end{equation}
where $\alpha\sim 1/137$ is the fine structure constant and $r$ is the typical
size of the heavy baryons.
Take $r\sim\Lambda_{\rm QCD}^{-1}=(300$ MeV$)^{-1}$ and $|\vec p_\gamma|\sim
300$ MeV, we have the estimate
\begin{equation}
\Gamma(\Lambda^{**}_Q\to\Lambda_Q\gamma)\sim 2 \hbox{ MeV}.
\end{equation}
Compare with the estimates of Ref.~\cite{13},
\begin{mathletters}
\begin{equation}
\Gamma(\Lambda_c(2593)\to\Sigma_c\pi)\sim 10 \hbox{ MeV},
\end{equation}
\begin{equation}
\Gamma(\Lambda_c(2625)\to\Lambda_c\pi\pi)\sim 0.1 \hbox{ MeV},
\end{equation}
\end{mathletters}
it is evident that the E1 decays are very probably observable, especially for
the $s={3\over2}$ state.
Hence it is of interest to calculate the electric dipole moments in the bound
state picture.

In the bound state model, $\Lambda^{**}_c$'s are bound states of heavy mesons
$D^{(*)}$ to nucleons N in the first excited state of a simple harmonic
potential \cite{8}.
\begin{equation}
|\Lambda^{**}_c\rangle = \textstyle{1\over\sqrt{2}}(|D^{(*)+}n^0\rangle +
|D^{(*)0}p^+\rangle)
\end{equation}
Denoting the position of the chiral soliton and the heavy meson as $\vec r_1$
and $\vec r_2$, and $\vec r=\vec r_1-\vec r_2$ their relative position, the
electric dipole moment can be written as,
\begin{equation}
\vec{\bf P}=\sum_{k=1,2}\int d^3r\;\phi_{1s}(\vec r)\; q_k \vec r_k \;
\phi_{2p}(\vec r),
\end{equation}
where the $\phi(\vec r)$'s the the simple harmonic wave functions, and the
$q_k$ are the electric charges carried by the chiral soliton and the heavy
meson respectively.
For the $D^{(*)+}-n^0$ system, $q_1=0$, $q_2=e$ and the dipole moment is
\begin{eqnarray}
\vec{\bf P}&=&e \int d^3r\;\phi_{1s}(\vec r)\; \vec r_2\; \phi_{2p}(\vec r)
\nonumber\\&=&-e{M_N\over M_N+M_D} \int d^3\;\phi_{1s}(\vec r)\;\vec r\;
\phi_{2p}(\vec r),
\end{eqnarray}
and for the $D^{(*)0}-p^+$ system,
\begin{eqnarray}
\vec{\bf P}&=&e \int d^3r\;\phi_{1s}(\vec r)\; \vec r_1\; \phi_{2p}(\vec r)
\nonumber\\&=&e{M_D\over M_N+M_D} \int d^3\;\phi_{1s}(\vec r)\;\vec r \;
\phi_{2p}(\vec r).
\end{eqnarray}
Hence, with
\begin{equation}
r_c=|\vec r_c|=|\langle 1s|\vec r|2p \rangle|=|\int d^3r\;\phi_{1s}(\vec r)\;
\vec r\; \phi_{2p}(\vec r)|,
\end{equation}
we sum up the two contributions above and get,
\begin{equation}
|\vec{\bf P}|={er_c\over 2}{M_D-M_N\over M_D+M_N}.
\end{equation}
Note the there is a cancellation between the $D^{(*)+}-n^0$ and the
$D^{(*)0}-p^+$ contribution.
Physically it is because the electric dipole vector is pointing from the
soliton to the heavy meson in the former case, and is reversed in the latter.
The cancellation is exact if $M_D=M_N$.
In the real world, $M_D/M_N\sim 2$ and this kinematical suppession factor is
about ${1\over3}$.

It is necessary to extract the parameters of the simple harmonic binding
potential to calculate $r_c$.
The reduced mass of the system is,
\begin{equation}
\mu_c=\left({1\over M_N}+{1\over M_D}\right)^{-1}=625 \hbox{ MeV}
\end{equation}
The spin-averaged excitation energy is,
\begin{equation}
\omega_c=\textstyle{1\over6}(4(340)+2(308)) \hbox{ MeV}=329 \hbox{ MeV}.
\end{equation}
Hence
\begin{equation}
r_c=\langle 1s|r|2p \rangle = (2\mu_c\omega_c)^{-1/2}=(641 \hbox{ MeV})^{-1},
\end{equation}
Putting all the pieces together, we have
\begin{mathletters}
\begin{equation}
\Gamma(\Lambda_c(2593)\to\Lambda_c\gamma)=0.016 \hbox{ MeV},
\end{equation}
\begin{equation}
\Gamma(\Lambda_2(2625)\to\Lambda_c\gamma)=0.021 \hbox{ MeV},
\end{equation}
\end{mathletters}
The small width is mainly due to the kinematical suppression factor.
Given the small width, it would be quite difficult to observe this radiative
decay for $\Lambda_c(2593)$.
On the other hand, the radiative $\Lambda_c(2625)$, may have a significant
branching ratio.

The electric dipole moment for the $\Lambda^{**}_b\to \Lambda_b\gamma$ decay
do not suffer the same kinematic suppression.
The $\Lambda^{**}_b$ is a linear combination of the $B^{(*)0}n^0$ bound state
and the $B^{(*)-}p^+$ bound state.
\begin{equation}
|\Lambda^{**}_b\rangle = \textstyle{1\over\sqrt{2}}(|B^{(*)0}n^0\rangle +
|B^{(*)-}p^+\rangle)
\end{equation}
The electric dipole moment of the $B^{(*)0}n^0$ system vanishes identically as
both the constituents are electrically neutral, while the $B^{(*)-}p^+$ system
has a nonvanishing electric dipole moment.

To calculate this electric dipole moment, it is again necessary to extract
the parameters of the simple harmonic binding potential.
The reduced mass $\mu_b$ for the $\Lambda^{**}_b$ system is
\begin{equation}
\mu_b=\left({1\over M_N}+{1\over M_B}\right)^{-1}=796 \hbox{ MeV}.
\end{equation}
One can predict the frequency of the harmonic potential $\omega_b$, which is
just the excitation energies of $\Lambda^{**}_b$ over the ground state
$\Lambda_b$, from the counterpart in the charmed system.
In Ref.~\cite{8}, it was shown that
\begin{equation}
{M_{\Lambda^{**}_b}\over M_{\Lambda^{**}_c}}=1-{1\over2}\left({M_N\over M_D}
-{M_N\over M_B}\right),
\end{equation}
which predicts the masses of the excited $\Lambda_b$'s at $\Lambda_b(5900)$
and $\Lambda_b(5926)$ (258 and 285 MeV above $\Lambda_b$ respectively).
Hence
\begin{equation}
\omega_b=\textstyle{1\over6}(4(285)+2(258)) \hbox{ MeV}=276 \hbox{ MeV}.
\end{equation}
As a result,
\begin{equation}
r_b=\langle 1s|r|2p \rangle = (2\mu_b\omega_b)^{-1/2}=(662 \hbox{ MeV})^{-1},
\end{equation}

The electric dipole moment is
\begin{eqnarray}
\vec{\bf P}&=&\textstyle{1\over2}\sum_{k=1,2}\int d^3r\;\phi_{1s}(\vec r)\;
q_k \vec r_k \;\phi_{2p}(\vec r)\nonumber\\
&=&\textstyle{e\over2}\int d^3r\;\phi_{1s}(\vec r)\; ({M_B\over M_B+M_N}
\vec r - {-M_N\over M_B+M_N}\vec r)\;\phi_{2p}(\vec r)\nonumber\\
&=&\textstyle{e\over2}\int d^3r\;\phi_{1s}(\vec r)\;\vec r\;\phi_{2p}(\vec r)\
=\textstyle{e\over2} \vec r_b,
\end{eqnarray}
which is expected as the electric dipole moment of a neutral system.
Finally putting all the pieces together, we end up with
\begin{mathletters}
\begin{equation}
\Gamma(\Lambda_b(5900)\to\Lambda_b\gamma)=0.090 \hbox{ MeV},
\end{equation}
\begin{equation}
\Gamma(\Lambda_b(5926)\to\Lambda_b\gamma)=0.119 \hbox{ MeV},
\end{equation}
\end{mathletters}
which is still smaller than the naive estimate above (essentially because the
quarks are actually fractionally charged which effectively weakens the
electromagnetic coupling) but yet measurable, and may even dominate for
$\Lambda_b(5926)$.
It is expected that this width (or equivalently the branching ratio) will be
measured within the next several years.

It should also be noted that, in this picture, the isospin allowed
$\Lambda^{**}_Q\to\Sigma_Q\gamma$ modes do not happen in leading order.
Physically it means that electric dipole transitions do not couple to spins
of particles.
Since $\Lambda$-type and $\Sigma$-type heavy baryons have different spin
structures (light valence quark spins are in a singlet state in the former
and in a triplet state in the latter case), the electric dipole between them
vanish identically.
Thus this mode must goes through a magnetic quadruapole transition, which is
severely suppressed.

The formalism described in this article can also be applied to the electric
dipole transitions of other orbitally excited heavy baryons.
For example, the orbitally excited $\Sigma^{(*)}_c$'s can decay to
$\Sigma^{(*)}_c\gamma$.
In the bound state picture, the $\Sigma^{(*)}_c$ baryons are linear
combinations of $D^{(*)}-$N and $D^{(*)}-\Delta$ bound states.
The calculations of the relevant electric dipole moments are similar in this
case, except one has to take the finite $\Delta-$N mass different into
account.
These orbitally excited $\Sigma^{(*)}_c$'s, however, can decay strongly
to the ground state $\Lambda_c$ and $\Sigma^{(*)}_c$ and these decays are
not phase space suppressed.
Hence the electromagnetic decays are expected to be unimportant in these
cases.

In conclusion, we have shown that the bound state picture predicts that the
$\Lambda^{**}_c\to \Lambda_c\gamma$ decay is severely suppressed.
On the other hand, the $\Lambda^{**}_b\to \Lambda_b\gamma$ is unsuppressed and
may have a significant branching ratio, especially for the $s={3\over2}$
state.
The $\Lambda^{**}_Q\to\Sigma_Q\gamma$ decay mode is expected to be small.
It is hopeful that these statements will be tested by experiments in the
coming years.

\acknowledgements
I would like to thank P. Cho and T.M. Yan for stimulating discussions.

\end{document}